\documentclass[reprint,showpacs,prd,superscriptaddress,floatfix]{revtex4-1}
\usepackage{dcolumn}
\usepackage{amsmath}
\usepackage{epsfig}
\usepackage{multirow}
\usepackage{epstopdf}

\def\UTfit{UT$_{\it fit}\ $}
\def\UTfitp{UT$_{\it fit}$}
\def\CP{$ C \! P$ }

\def\ra{\rightarrow}

\def\sbabar{\mbox{{\normalsize \sl B}\hspace{-0.4em} {\small \sl A}\hspace{-0.03em}{\normalsize \sl B}\hspace{-0.4em} {\small \sl A\hspace{-0.02em}R}}}
\let\babar=\sbabar

\newcommand\T{\rule{0pt}{2.6ex}}       
\newcommand\TT{\rule{0pt}{3.0ex}}       
\newcommand\B{\rule[-1.2ex]{0pt}{0pt}} 

\begin{document}

\hfill CALT 68-2902

\title{Global CKM Fits with the Scan Method}
\author{Gerald Eigen} 
\affiliation{University of Bergen, 5007 Bergen, Norway}
\author{Gregory Dubois-Felsmann}
\affiliation{SLAC National Accelerator Laboratory, Stanford, California 94309, USA}
\author{David G. Hitlin}
\author{Frank C. Porter}
\affiliation{California Institute of Technology, Pasadena, California 91125, USA}

\vskip3cm

\begin{abstract}


We present results of unitary triangle fits based on the scan method. This frequentist approach employs Gaussian uncertainties for experimental quantities, but avoids assumptions about the distribution of theoretical errors. Instead, we perform a large number of fits, scanning over regions of plausible theory errors for each quantity. We retain those fits meeting a specific confidence level criterion, thereby constructing a region in the $\bar \rho - \bar \eta $ plane using the ``standard'' measurements (CKM matrix elements, $\sin2 \beta, B^0_{d,s}$ mixing, $\epsilon_K$).  In addition we use branching fraction and \CP asymmetry measurements of $B$ decays to pseudoscalar pseudoscalar, 
pseudoscalar vector, vector vector and $a_1$ pseudoscalar
 final states to determine $\alpha$, $D^{(*)}K^{(*)}$ modes to determine $\gamma$, and $D^{(*)}\pi$ and $D\rho$ modes to determine $2\beta +\gamma$. We parameterize individual decay amplitudes in terms of color-allowed tree, color-suppressed tree, penguin, singlet penguin, electroweak penguin, 
 as well as  $W$-exchange and $W$-annihilation amplitudes. With this parameterization, we obtain a good fit to the measured branching fractions and \CP asymmetries within the Standard Model {\it ansatz}, with no new physics contributions. This simultaneous fit allows us to determine, for the first time in a global fit, the correlation between $\alpha$ and $\beta$, as well as between $\gamma$ and $\beta$.

\end{abstract}

\maketitle

\section{Introduction}

The phase of the Cabibbo-Kobayashi-Maskawa (CKM) matrix~\cite{ckm} is responsible for \CP violation in the Standard Model (SM). The unitarity relations within the CKM matrix provide an excellent laboratory to test this prediction, the relation ${V_{ub}}^* V_{ud} +{V_{cb}}^* V_{cd} +{V_{tb}}^* V_{td} =0$ being particularly useful, since many measurements and theory inputs in the $B $ and $K$ systems can be combined for this test.  Tests (including 
simultaneous estimation of the fundamental CKM parameters) such as those by the CKMfitter~\cite{ckmfitter} and \UTfit\cite{utfit} groups are in common circulation. The former is a frequentist technique, the latter Bayesian. The scan method~\cite{scan} presented herein is a frequentist-based fitting technique to determine the parameters of the CKM matrix and test
consistency with the standard model.  The scan method takes a different approach in its treatment of theoretical uncertainties, as well as in the construction of confidence sets.

We first describe the scan method in section~\ref{sec:fitMeth}. The results for the global fits
are presented in section~\ref{sec:globalFits}. We begin with a comparison with  CKMfitter and \UTfitp, 
 employing the scan method to extract CKM parameters using inputs standardized for the book {\sl Physics of the $B$ Factories}~\cite{pbf}.
Then we look at the current situation using inputs from PDG12~\cite{pdg12} and HFAG~\cite{HFAG},  investigating the question of consistency of the results with SM expectations. We call such fits ``baseline" fits. Their characteristic is that the $\chi^2$ function has of the order of ${\cal O}(20)$ terms and ${\cal O}(10)$ fit parameters including explicit  inputs for $\alpha$ and $\gamma$. To take into account the correlations between $\alpha$ and $\beta$ or $\gamma$ and $\beta$ in the extraction of the parameters of the unitarity triangle, we perform a fit in which we replace the inputs for  $\alpha$ and $\gamma$ by branching fractions and \CP asymmetries of $B$ decays to pseudoscalar pseudoscalar ($PP$), pseudoscalar vector ($PV$), vector vector ($VV$) and $a_1$ pseudoscalar ($a_1 P$) final states and $B$ decays to $D^{(*)}K^{(*)}$, $D^{(*)}\pi$, and  $D\rho$ final states. We refer to the latter as ``full" fits. Thus, we are able for the first time to include the correlations between $\alpha$ and $\beta$ and $\gamma$ and $\beta$ in the extraction of the parameters of the unitarity triangle. 
 
In section~\ref{sec:alphaGamma}, we perform stand-alone determinations of the angles $\alpha$ and $\gamma$, which are then used in the baseline fits.
First, we determine the correlations between $\alpha$ and $\beta$ by performing fits to measurements of $B$ decays to $PP, PV, VV$, and $a_1 P$ final states. Next, we determine the correlations between $\gamma$ and $\beta$ by performing fits to measurements of $B$ decays to $D^{(*)}K^{(*)}$, $D^{(*)}\pi$ and $D\rho$ final states.
Finally, 
the results are summarized in section~\ref{sec:conclusion}.

\vfill\break

\section{Fit Methodology}
\label{sec:fitMeth}

The scan method accounts for the theoretical uncertainties in the QCD parameters $f_{B_s}$, $\xi_f=f_{B_s}/f_{B_d}$, $B_{B_s}$, $\xi_b=B_{B_s}/B_{B_d}$, and $B_K$~\cite{laiho10, laiho2} and the CKM parameters $|V_{ub}|$ and $|V_{cb}|$ by scanning over the range allowed by theory uncertainties using fixed grid or Monte Carlo (MC) methods. In the baseline fit, we combine measurements of $\Delta m_{B_d}, \Delta m_{B_s}, \epsilon_K, |V_{cb}|, |V_{ub}|, |V_{ud}|$, $|V_{us}|$, $|V_{cd}|$, $|V_{cs}|$, $|V_{tb}|$, $\sin2 \beta, \alpha$ and $\gamma$ in the $\chi^2$ function, Eq.~\ref{eq:chisq}. The angle
brackets ($< >$) here indicate the experimental averages.

\begin{widetext}

{\large
\begin{eqnarray}
\label{eq:chisq}
\chi^2(\bar \rho, \bar \eta, p_i, t_j)=
&\Bigl (\frac{\langle  \Delta m_B{_{d,s}}        \rangle -\Delta m_{B_{d,s}}(\bar \rho, \bar \eta, p_i, t_j )} {\sigma_{\Delta m_{B_{d,s}}}}   \Bigr )^2  +
\Bigl ( \frac{\langle | V_{cb,ub,ud,us} |  \rangle - |V_{cb,ub,ud,us}| (\bar \rho, \bar \eta, p_i, t_j ) } {\sigma_{|V_{cb,ub,ud,us}|}}   \Bigr)^2 \nonumber \\
+&\Bigl ( \frac{\langle | \epsilon_K |           \rangle - \epsilon_K  (\bar \rho, \bar \eta, p_i, t_j)} {\sigma_{\epsilon_K}}                 \Bigr)^2 +
\Bigl ( \frac{\langle S_{\psi K^0}            \rangle - \sin2 \beta (\bar \rho, \bar \eta, p_i)} {\sigma_{S_{\psi K^0}}}                        \Bigr)^2 +
\Bigl ( \frac{\langle \alpha                       \rangle - \alpha(\bar \rho, \bar \eta, p_i)} {\sigma_\alpha}                                                   \Bigr)^2  \nonumber \\
+&\Bigl ( \frac{\langle \gamma                    \rangle - \gamma(\bar \rho, \bar \eta, p_i)} {\sigma_\gamma}                                              \Bigr)^2 +
\sum_k \Bigl ( \frac{\langle {\cal M}_k   \rangle - {\cal M}_k (p_i)} {\sigma_{{\cal M}_k}}                                                                    \Bigr)^2 +
\sum_n  \Bigl ( \frac{\langle {\cal T}_n  \rangle - {\cal T}_n(p_i, t_j)} {\sigma_{{\cal T}_n}}                                                              \Bigr)^2.
\end{eqnarray}
}

\end{widetext}

\begin{table*}[ht]
\centering
\caption{Observables used in the baseline fits and full fits. Fit type I are baseline fits with inputs specified by CKMfitter and \UTfitp\ (July 2012) to compare the fit
results among the different fit methodologies. The values of $|V_{cb}|$  and $ |V_{ub}| $ have only a total uncertainty. We list a second set in which experimental and theory uncertainties are separated allowing scans over the theoretical component of the uncertainties in $|V_{cb}|$  and $ |V_{ub}| $.  
Fit type II are the baseline fits using the most recent input values to test the SM with and without the inclusion of ${\cal B}(B^+  \ra \tau^+ \nu)$. 
The first set of $\alpha$ and $\gamma$ values has been fixed by CKMfitter and \UTfitp, while the second set results from our global fits to branching fractions and \CP\  asymmetries of $B$ decay modes.
Fit type III represents the full fits in which $\alpha$ and $\gamma$ are replaced by branching fraction and \CP asymmetry measurements of the $B$ decay modes.}
\medskip

{
\begin{tabular}{|c|c|c|c|c|}
 \hline \hline 
fit type &$m_t ^{\rm pole}~ \rm [GeV/c^2] $ & $\overline{m}_c(m_c)~ \rm [GeV/c^2] $& $\Delta m_{B_d} ~ \rm [ps^{-1}]$ & $\Delta m_{B_s} ~ \rm [ps^{-1}]$ \T \B \\ \hline
I& $173.2 \pm 0.9$~\cite{pbf} & $1.275 \pm 0.025$~\cite{pdg12} & $0.508 \pm 0.004$~\cite{pdg12} & $ 17.719 \pm 0.042$~\cite{pdg12}  \T \B \\ 
II, III & $173.5 \pm 0.9$~\cite{pdg12} & same & same & same \T \B \\ \hline \hline
fit type & $|V_{cb} |$ & $| V_{ub} |$ & $ |V_{ud}| $ & $| V_{us}| $ \T \B \\ \hline
I & $(4.16 \pm 0.038\pm0.05) \times 10^{-2} $~\cite{pbf} & $(3.95 \pm 0.38\pm0.39) \times 10^{-3}$~\cite{pbf} & $\ 0.97425 \pm 0.0002\ $~\cite{colangelo}   & $\ 0.2208\pm 0.0039$~\cite{colangelo}  \T \B \\ 
II, III & $(4.09 \pm 0.07\pm 0.09) \times 10^{-2}  $~\cite{pdg12}&$(4.15 \pm 0.31 \pm 0.39) \times 10^{-3}   $~\cite{pdg12} & same  &  same  \T \B \\ \hline \hline
fit type & $ \epsilon_K $ & $ \sin 2 \beta$ & $ \alpha $ & $ \gamma$ \T \B \\ \hline
I  & $(2.228\pm 0.0011) \times 10^{-3}$~\cite{pdg12} & $ 0.0677\pm 0.020$~\cite{HFAG} &$ (88.0\pm   
5.0)^\circ$~\cite{pbf}  & $ (67.0\pm 11)^\circ$~\cite{pbf} \T \B \\  
II, III & same &  same  & $ (84.6\pm 2.1)^\circ$~\footnote[1]{Input from global fit to Unitarity Triangle angles (see below); not used in the full fit (III). } & $ (79.7\pm 4.2)^\circ$${\rm \ ^a}$ \T \B \\ 
\hline \hline      
fit type & $|V_{cd}|$ & $|V_{cs}|$  & $|V_{tb}|$ & ${\cal B}(B^+ \rightarrow \tau^+ \nu) $ \T\B \\ \hline
I  & not used & not used & not used &  $(1.15\pm0.23) \times 10^{-4}$    \T\B \\
II, III & $0.23 \pm 0.011$~\cite{pdg12} &   $ 1.023\pm 0.036$~\cite{pdg12} & $0.97 \pm 0.08$~\cite{pdg12} & same \T\B \\       
\hline \hline      
\end{tabular}
}
\label{tab:observables}
\end{table*}

\begin{table*}[ht]
\centering
\caption{QCD parameters used in the baseline fits. The first row shows the inputs for fit type I that  were the averages listed by the Lattice group in July 2012. The second row shows the values with separate ``statistical" and theory uncertainties used in the baseline fit II and the full fit (III). }\medskip
{
\begin{tabular}{|c|c|c|c|c|c|c|}
\hline \hline
fit type & $f_{B_s}$ [MeV] & $\xi_f  $ & $B_{B_s} $  & $\xi_b $  & $B_K$& Ref. \T \B  \\ \hline
I & $\ 227.6 \pm 2.2 \pm 4.5$  &  $\ 1.201 \pm 0.012 \pm 0.012$   & $\ 1.33\pm 0.06$ & $1.05 \pm 0.07$  & $\ 0.7643\pm 0.0034\pm 0.0091$&\cite{laiho2, laiho10} \T \B \\ 
II, III & same &  same   & $\ 1.33\pm 0.018\pm 0.06$ & $1.05 \pm 0.025\pm 0.07$  & same &\cite{gamiz} \T \B \\ 
\hline \hline

fit type & $\eta_{cc}$ &  $\eta_{ct}$ &  $\eta_{tt}$ &  $\eta_{b}$ & &\T \B \\ \hline
I, II, III& $\ 1.39 \pm 0.35$ & $ \ 0.47 \pm 0.04$ & $\ 0.5765 \pm 0.0065$ & $\ 0.551\pm 0.007$ & &\cite{nierste} \T \B \\
\hline \hline
\end{tabular}
}
\label{tab:qcd}
\end{table*}

\noindent
The dependence of the predicted values on the quantities $\bar \rho, \bar \eta, p_i$, and $t_j$
is described in detail in the Appendix. The $p_i$ are measured  inputs to these predicted values, including the Wolfenstein parameters~\cite{wolf} $A$ and $\lambda$ and the quark masses and $B$ meson masses.   We add terms in the $\chi^2$, denoted generically by ${\cal M}_k$, accounting for the contributions from
the uncertainties in the $p_i$. Note that the dependence on these terms introduces
correlations in the $\chi^2$ expression. 
The $t_j$ represent parameters having a theoretical uncertainty, {\it e.g.} the QCD parameters as well
as $|V_{ub}|$ and $|V_{cb}|$.
 
We are careful to distinguish among different kinds of uncertainties. Observables with experimental errors only (statistical and systematic) are assumed to be Gaussian-distributed. Theoretical quantities such as lattice-derived QCD parameters, and inputs to $|V_{ub}|$ and $|V_{cb}|$, typically have two types of uncertainties.
The first type of error is of a ``statistical" nature, resulting from an input with statistical uncertainties or from Monte Carlo statistics in lattice calculations. We assume this error to be Gaussian-distributed and add corresponding terms to the $\chi^2$, 
denoted by ${\cal T}_n$. The second type of uncertainty is a theory error with no known underlying statistical distribution. We therefore make no assumption as to the distribution of these errors, and instead perform a scan over a large range of plausible values, doing a $\chi^2$ minimization at each point. Our use of such fit results is described in the next section.

The scan includes the QCD parameters ($f_{B_s}$, $\xi_f$, $B_{B_s}$, $\xi_b$ and $B_K$) and the CKM matrix elements $|V_{ub}|$ and $|V_{cb}|$. The QCD corrections $\eta_{cc}, \ \eta_{ct}$ and $\eta_{tt}$ used in the determination of $\epsilon_K$ and $\eta_b$ that appear in the prediction of $\Delta m_{B_d}$ also have theory errors. 
Although the fit methodology is able to scan over them, we do not do so, since this is unnecessary at the current level of precision. We parametrize $\eta_{cc}$ in terms of $\overline{m}_c(m_c)$ and $\alpha_s$~\cite{nierste}. Tables~\ref{tab:observables} and \ref{tab:qcd} summarize the input parameters for our baseline fits. We are preparing a more detailed article that includes a fuller discussion of all input values, provides a study of the correlations among the theory uncertainties and shows further results~\cite{unfit}.

\section{Results of the Global Fits}
\label{sec:globalFits}

We present in this section the results for the global fits according to the
scan method, beginning with a comparison with the results for CKMfitter and \UTfit\ for
a common set of inputs. Then we look at the scan method results for a more current 
set of inputs, including both a stand-alone determination of the angles $\alpha$
and $\gamma$ and a fit explicitly incorporating the measurements that enter into the
$\alpha$ and $\gamma$ determination.

\subsection{Comparison to CKMfitter and \UTfit}

\begin{figure*}[htbp!]
\begin{minipage}[b]{0.48\textwidth}
\centering
\includegraphics[width=\textwidth, height=85mm]{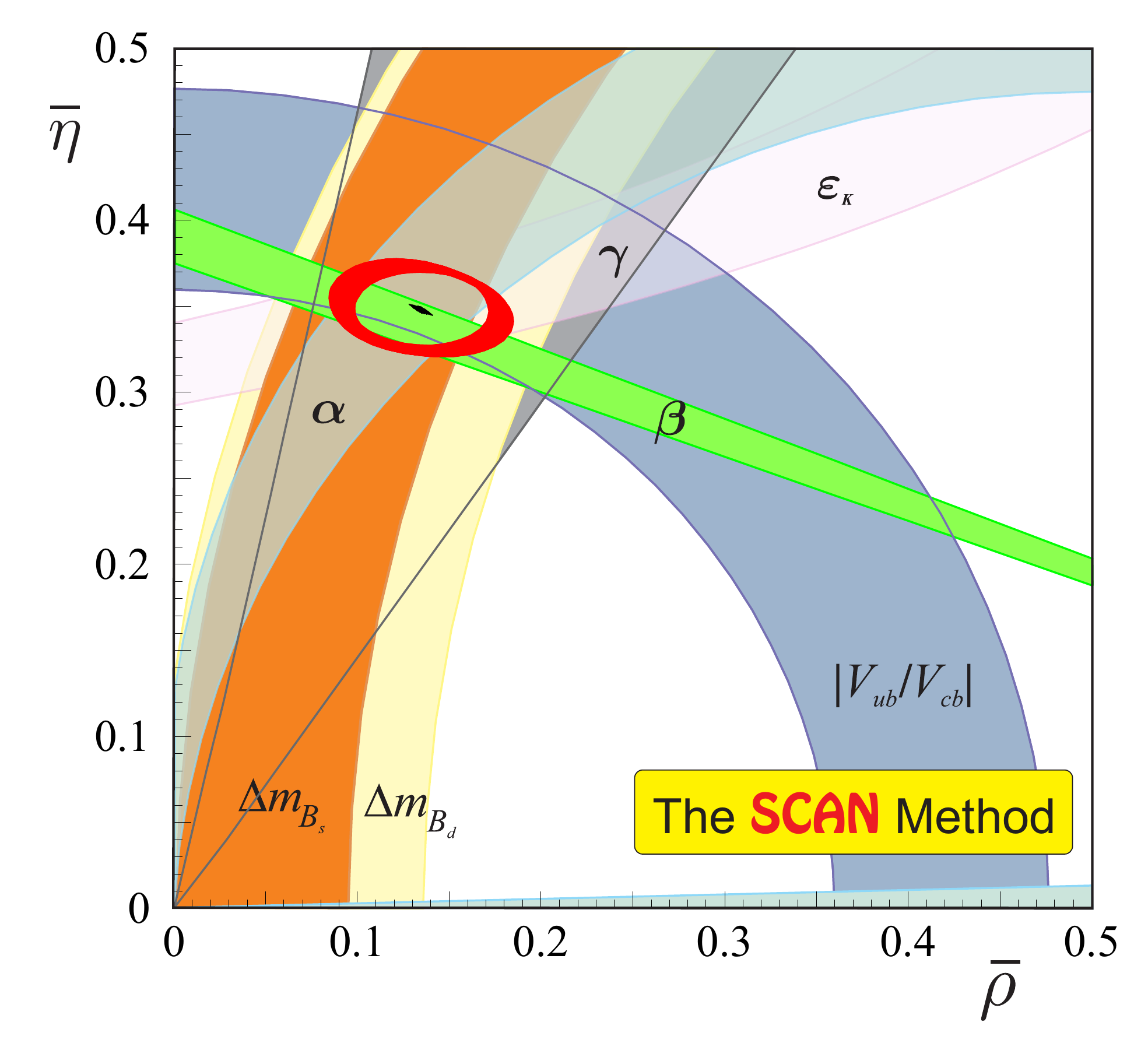}
\vskip -12pt
\caption{Overlay of $68\%~{\rm CL}$ contours in the $\bar \rho -\bar \eta$ plane for the inputs of {\sl Physics of the $B$ Factories} with scanning over $f_{B_s}$, $\xi_f$ and $B_K$. The black points show the central values of each accepted fit. 
}
\label{fig:pbf}
\end{minipage}\quad
\begin{minipage}[b]{0.48\textwidth}
\centering
\includegraphics[width=\textwidth, height=85mm]{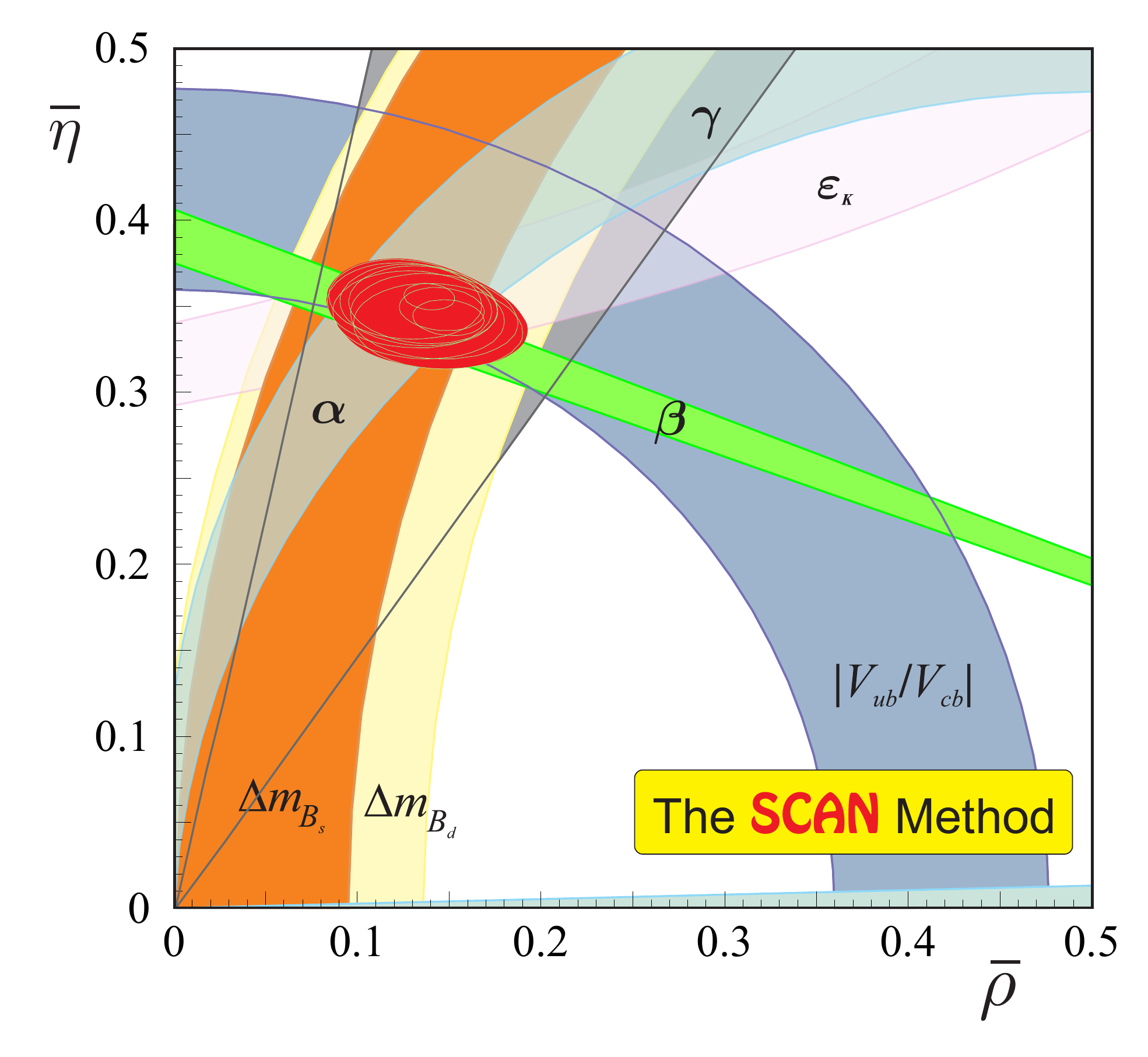}
\vskip -12pt
\caption{Overlay of $68\%~{\rm CL}$ contours in the $\bar \rho -\bar \eta$ plane for the inputs of {\sl Physics of $B$ the Factories} with  scanning over $f_{B_s}, \xi_f, B_K, |V_{ub}|$ and $|V_{cb}|$. The green ellipses show selected contours of accepted fits.}
\label{fig:pbf2}
\end{minipage}
\end{figure*}

\begin{table*}[ht!]
\centering
\caption{Comparison of unitarity triangle parameters for different fitting techniques using inputs for the book {\sl Physics of the $B$ Factories}.  The second and third columns show the fit results from CKM fitter and \UTfitp.  The fourth column shows our fit result if no scanning over theory parameters is performed and experimental uncertainties and theory uncertainties are added in quadrature. The fifth column shows the fit results if we scan over the QCD parameters  $f_{B_s}, \xi_f {\rm and\ } B_K$. The sixth column shows the fit results if we scan, in addition,  over the theory errors associated with $|V_{ub}|$ and $|V_{cb}|$.}\medskip
{
\begin{tabular}{|l||c|c|c||c|c|}
\hline \hline
  & &  & Our fit & \multicolumn{2}{c|}{Scan method$\ \ \ \ \ \ \ \ \ \ \ \  $} \T \B   \\
 Parameter & CKMfitter &\UTfit & no scan & scan over  & scan over    \\ 
& \cite{pbf} & \cite{pbf} & & $\ f_{B_s}, \xi_f, B_K\ $ & $\ f_{B_s}, \xi_f, B_K$,  $|V_{cb}|$, $|V_{ub}| \ $  \B  \\ \hline\hline
$\ \bar \rho $ &  $\  0.129^{+0.027}_{-0.022}$ & $\ 0.132\pm 0.020\  $& $\ 0.134\pm 0.041\ $ & $ 0.132^{+0.048}_{-0.042}$ & $0.139^{+0.048}_{-0.052} $  \TT \B \\
$\ \bar \eta $ &  $\  0.345\pm 0.014\ $ & $\ 0.348\pm 0.013$ & $\ 0.348^{+0.024}_{-0.023}$ & $0.348^{+0.026}_{-0.025}$  & $ 0.341^{+0.034}_{-0.025}$   \T \B \\
$\ \beta~[^\circ] $ & $ 21.6^{+0.8}_{-0.7}$ & $ 21.8^{+0.8}_{-0.7}$ & $\ 21.9^{+2.4}_{-2.2}$ & $21.8^{+2.5}_{-2.1}$ & $ 21.6^{+2.6}_{-2.2}$ \T \B  \\
$\ \alpha~[^\circ]$ & $ 88.8^{+4.2}_{-3.6} $ & $ 88.6\pm 2.9 $ &$\ 89.2^{+1.7}_{-2.9} $& $ 89.0^{+2.9}_{-3.6}$ & $ 90.6^{+2.8}_{-5.9}$  \T \B  \\
$\ \gamma~[^\circ]$ & $ 68.96^{+3.5}_{-4.2} $&$ 69.4\pm 3.1 $ &\ $ 68.9^{+5.1}_{-4.1}$ & $69.2^{+5.5}_{-5.2}$&$ 67.8^{+7.4}_{-5.1}$ \T \B  \\

\hline \hline
\end{tabular}
}
\label{tab:comparison}
\end{table*}

To begin, we compare the performance of the scan method with that of CKMfitter and
\UTfit using 19 input measurements ($|V_{ud}|$, $ |V_{us}|$,  $|V_{cb}|$,  $|V_{ub}|$, 
$\epsilon_K$, $\Delta m_{B_d}$,  $\Delta m_{B_s}$,  $\sin 2\beta$, $ \alpha$, $\gamma$,  $f_{B_s}$, $B_{B_s}$, $\xi_f$, $\xi_b$, $B_K$,  $m_t^{\rm pole}$,  $\overline{m}_c(m_c)$, $m_{B_d}$,  $m_{B_s}$), choosing values specified for the book {\sl Physics of the $B$ Factories}~\cite{pbf} to fit 13 parameters ($\bar \rho$, $\bar \eta$, $A$, $\lambda$, $f_{B_s}$, $B_{B_s}$, $\xi_f$, $\xi_b$, $B_K$, $m_t^{\rm pole}$, $\overline{m}_c(m_c)$, $m_{B_d}$, $m_{B_s}$). In these fits, which we call fit type I in Tables~\ref{tab:observables} and ~\ref{tab:qcd},
we use the central values and measurement errors for $\alpha$ and $\gamma$ in the $\chi^2$ function. We compute $\overline{m}_t(m_t)$, which enters into the Inami-Lim functions~\cite{inami} for $\Delta m_{B_d}$, $\Delta m_{B_s}$ and $\epsilon_K$, in the $\overline{MS}$ scheme from the pole mass $m_t^{\rm pole}$ at three loop level for six quarks~\cite{melnikov00, gray90, broadhurst91}.
We plot $1\sigma$ contours in the $\bar \rho- \bar \eta$ plane according to the 
prescription described below (see Eq.~\ref{Eq:testRegion}).
For the central value, we select the fit with the highest  $P(\chi^2)$. We take the $\pm1\sigma$ uncertainties from the maximum and minimum values of the envelope of all contours. We perform three different fits. In the first, we combine theory and experimental uncertainties, treating these as Gaussian. In the second, we scan over the theory uncertainties in $f_{B_s}, B_{B_s}, B_K$. In the third, we separate  theory uncertainties from experimental uncertainties in $|V_{ub}|$ and $|V_{cb}|$ (see Table~\ref{tab:observables}) thus scanning over theory uncertainties in $|V_{ub}|, |V_{cb}|$, as well as those for $f_{B_s}, \xi_f$ and $ B_K$.   We performed the separation according to the ratio of experimental to theory uncertainties listed in PDG10~\cite{pdg10} keeping the total uncertainty unchanged.
Figure~\ref{fig:pbf} shows the overlay of $68\%$ confidence level contours in the $\bar \rho - \bar \eta$ plane for the second fit. Since values of  $|V_{ub}|$ and $|V_{cb}|$, which have substantial theoretical uncertainties, are not scanned the accepted contours are similar. Figure~\ref{fig:pbf2} shows the comparable results for the third fit, {\it i.e}, the most complete scan.
Table~\ref{tab:comparison} lists our results in comparison to those from CKMfitter and \UTfitp.  The three methods yield broadly similar results.  
Scanning over
theoretical errors increases the uncertainties, often substantially.
With this methodology the resulting uncertainties are typically around a factor of two larger than those from CKMfitter and \UTfitp. 
Large differences may be expected with \UTfitp, as \UTfitp\ is a Bayesian approach and includes
prior distributions for the theoretically uncertain quantities. 

The differences with
CKMfitter are more subtle, as both are frequentist approaches. The essential difference between
the two interval estimation methods is the following: CKMfitter uses the conventional change in
$\chi^2$ from its minimum value to determine confidence regions. The motivation for the scan method is to test the hypothesis that the standard model is correct, against the alternative that it is incorrect. The statistical test adopted is the $\chi^2$ test. Hence,
this motivation is reflected in our confidence region determination, in which we use the method of
inverting a test acceptance region~(as described in standard statistics texts, for 
example~\cite{Shao2003}).  That is, the confidence regions are determined by comparing
$\chi^2$ values with a critical value instead of looking for a change in $\chi^2$.

Specifically, the algorithm for a $1-\alpha$ confidence region in $d$ dimensions of a $p$ dimensional parameter space with $n$ measurements is as follows. First, determine the 
acceptance region, at the $\alpha$ significance level, by determining the critical
value $\chi_c^2$ such that
\begin{equation}
 P(\chi^2\ge \chi_c^2; n-p+d|H_0) \ge \alpha,
 \label{Eq:testRegion}
\end{equation}
where $H_0$ is the hypothesis of the standard model, and $n-p+d$ is the number of degrees of freedom. The $d$ is
added back here because the critical region for the test is constructed for each point in the $d$-dimensional subspace of the full parameter space. The confidence region is then given by
all those points in the $d$ dimensional parameter subspace for which $\chi^2\le\chi^2_c$, under $H_0$.

For the theoretical uncertainties, CKMfitter makes an implicit scan by incorporating a term in the likelihood
function for each theoretically uncertain quantity. The term equals 1 or 0 depending on whether the theory parameter is in the theoretically ``allowed'' region or not. A modified algortihm is also available, which makes a smooth transition between
1 and 0. 
In the scan method, the scan over theoretical parameter space is explicit. No theoretical term is included in the
likelihood function. Instead, each point in the theoretical parameter space of interest is treated
as a possible value, and the fit procedure is performed at each such point.

Depending on the value of the minimum $\chi^2$, the regions determined by either
method may be larger or smaller. If the best fit gives a high $p$-value, then the
scan method regions will be larger. On the other hand, if the best fit gives a low $p$-value, the
CKMfitter regions will be larger. In the present instance, the best fit $p$-value is 0.76, hence the scan method confidence intervals will tend to be larger than those from CKMfitter.  Reflecting its origins in goodness-of-fit testing, the scan method region goes to null in the limit where the fit fails at the specified confidence level. In the limit of no theoretical uncertainties, both methods give
valid frequentist confidence regions, with different properties. However, this comparison points out the importance of
methodology in forming conclusions, and hence of examining the problem with multiple approaches.

\begin{table*}[ht]
\centering
\caption{The $95\%~{\rm CL}$ ranges for  unitarity triangle parameters from our baseline fit scans without the inclusion of ${\cal B} (B^+ \ra \tau^+ \nu)$, from our baseline fit scans with the inclusion of
${\cal B} (B^+ \ra \tau^+ \nu)$ and our full fit scans without the inclusion of ${\cal B} (B^+ \ra \tau^+ \nu)$. The values of $\alpha$ and $\gamma$ are, in these cases,  computed from the values of $\bar{\rho}$ and $\bar{\eta}$.} \medskip
{
\begin{tabular}{|l|c|c|c|c|c|}
\hline \hline
$\ \ $Parameter        & $\bar \rho$      &$\bar \eta$                        & $\beta~ [^\circ]$   & $\alpha~ [^\circ]$  & $\gamma~ [^\circ]$ \T \B \\ \hline
$\ \ $baseline fit: scan without $B^+ \ra \tau^+ \nu\ $    & $\ 0.069 -0.147\ $  & $\ 0.319-0.395\ $   & $\ 19.0-24.7\ $   & $\ 82.7-88.5\ $  & $\ 68.8-77.9\ $  \T \B  \\
$\ \ $baseline fit: scan with $B^+ \ra \tau^+ \nu$         & $ 0.073-0.147$  & $ 0.324-0.396$   & $ 19.3-24.8$   & $ 82.8-88.4$ & $ 68.8-77.4$\T \B  \\
\hline
$\ \ $full fit: scan without $B^+ \ra \tau^+ \nu$         & $ 0.070-0.151$  & $ 0.318-0.395$   & $ 18.8-24.8$   & $ 82.4-89.0$ & $ 67.9-77.9$\T \B  \\
 \hline \hline
\end{tabular}
 }
\label{tab:SMfit}
\end{table*}

\begin{figure}[h]
\begin{minipage}[b]{0.5\textwidth}
\centering
\includegraphics[width=90mm]{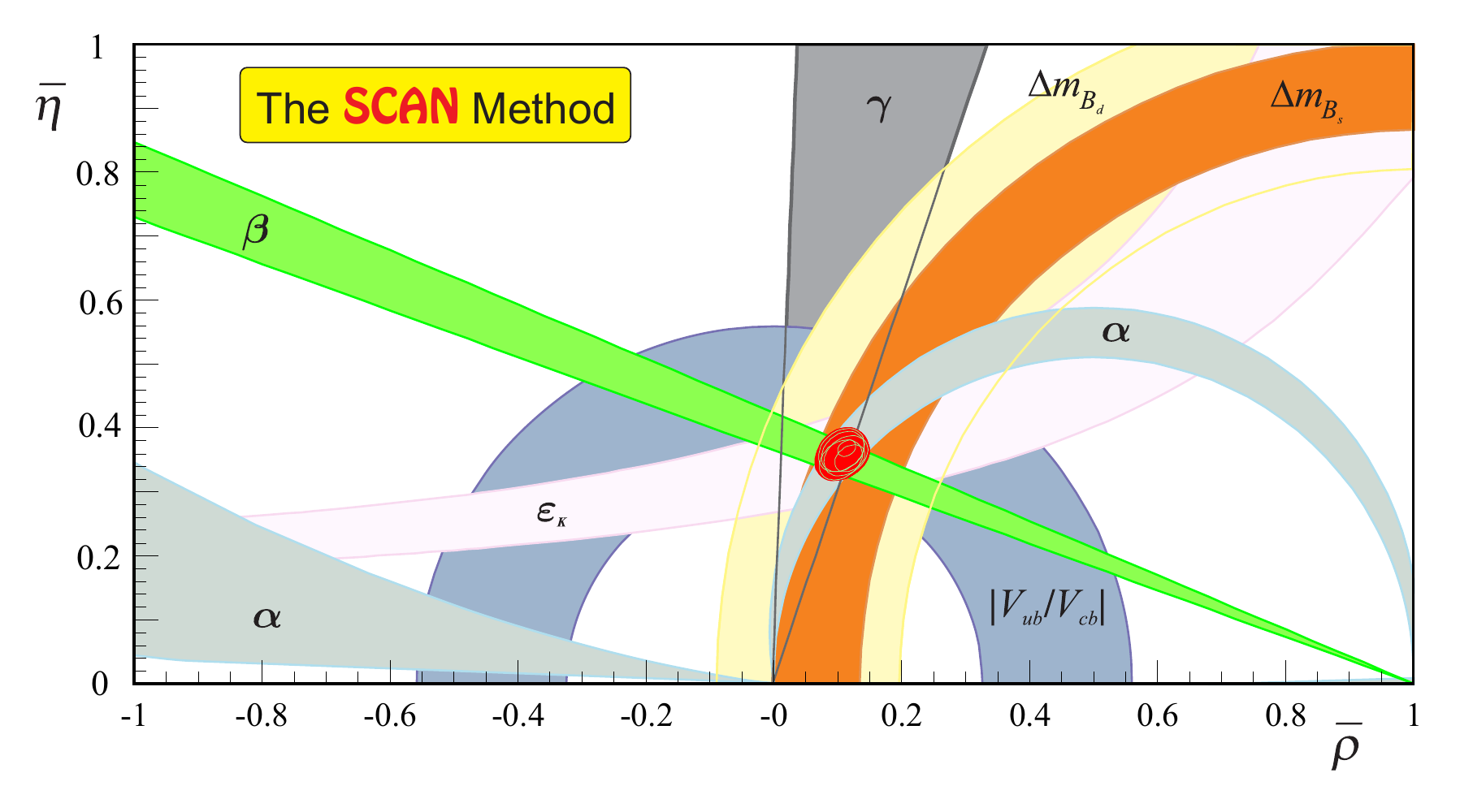}
\vskip -1pt
\caption{Overlay of $95\%~{\rm CL}$ contours in the $\bar \rho -\bar \eta$ plane for the baseline fit scan with 22~measurements without including ${\cal B}(B^+  \ra \tau^+ \nu)$. The green ellipses show selected contours of accepted fits.}
\label{fig:gfit}
\end{minipage}

\hspace{0.5cm}

\begin{minipage}[b]{0.5\textwidth}
\centering
\includegraphics[width=90mm]{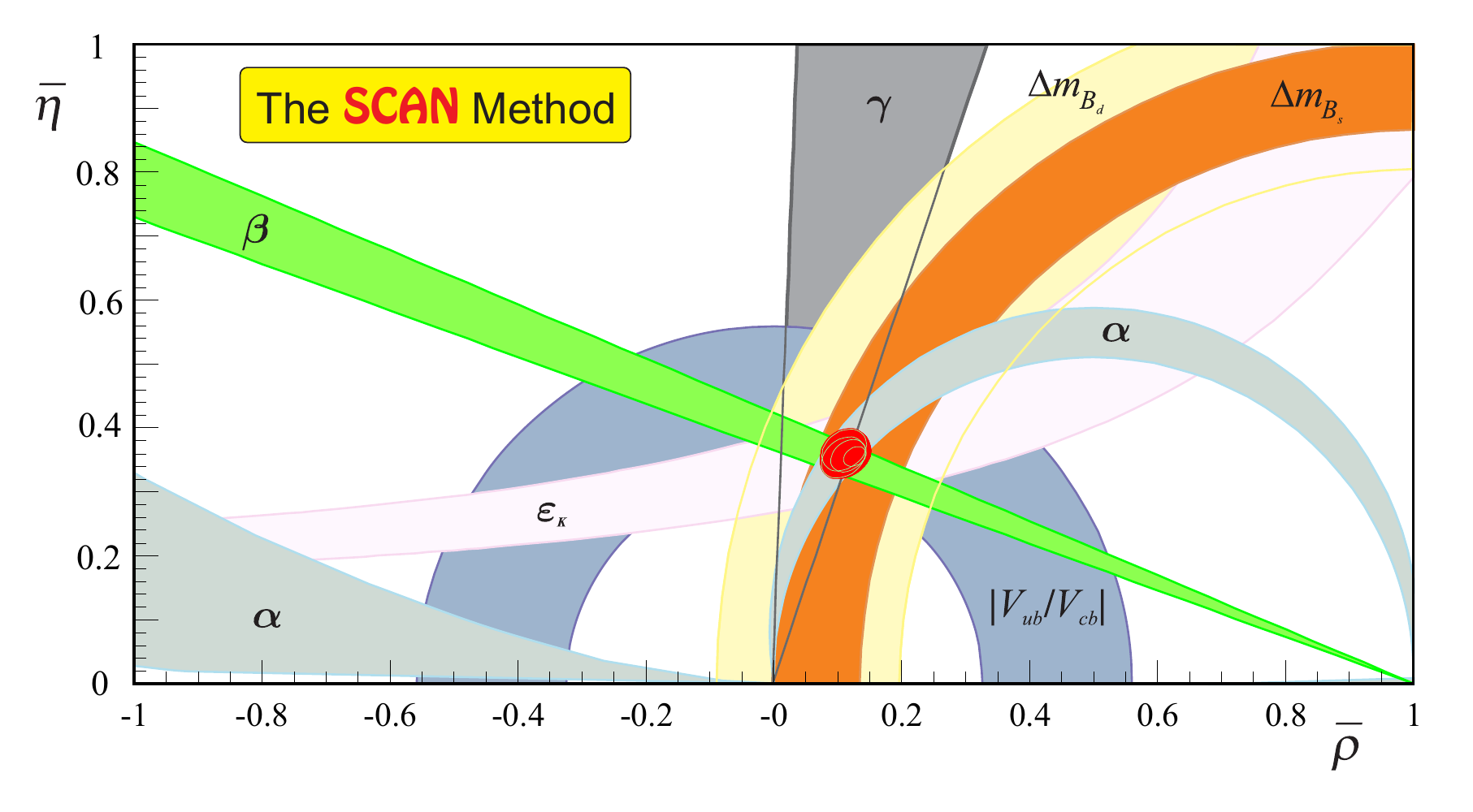}
\vskip -1pt
\caption{Overlay of $95\%~{\rm CL}$ contours in the $\bar \rho -\bar \eta$ plane for the baseline fit scan with 23~measurements with inclusion of ${\cal B}(B^+  \ra \tau^+ \nu)$. The green ellipses show selected contours of accepted fits.}
\vskip -18pt
\label{fig:gfit2}
\end{minipage}
\end{figure}

\subsection{Fit results with direct inputs of $\alpha$ and $\gamma$}

To test the SM with the scan method, we perform the same baseline fits with updated input parameters also
listed in Tables~\ref{tab:observables} and ~\ref{tab:qcd}. We refer to these fits as type II fits. We use  PDG12 $|V_{ub}|$ and $|V_{cb}|$  averages~\cite{pdg12}. Since the  $|V_{ub}|$ and $|V_{cb}|$  results from exclusive modes are significantly lower than those from inclusive modes, the PDG uses
scaling factors on the total errors of 2.6 and 2.0, respectively. 
If we have any 
fit satisfying $P(\chi^2)>5\%$, the SM is deemed to be compatible with the data at the present  level of theoretical
uncertainties.  
We compute 95\% CL contours according to the inversion of a test acceptance
region prescription described above. That is, we compute contours based on
Eq.~\ref{Eq:testRegion}, with d=1, such that taking the extrema of a contour along a parameter
axis provides a 95\% CL interval for the parameter.
In the limit of no
theoretical uncertainties, the procedure has the stated coverage. 
Figure~\ref{fig:gfit} shows the overlay of $95\%~{\rm CL}$ contours in the $\bar \rho-\bar \eta$ plane for all accepted baseline fits using 22 measurements to fit 13 parameters, as in Section~III. 
To obtain a range of values for a given parameter, we take the extrema of the union of the contours attached to accepted fits. Table~\ref{tab:SMfit} shows the range of the unitarity triangle parameters thus obtained (hereinafter referred to as the  $95\%~ {\rm CL}$ range). 

The $B^+ \ra \tau^+ \nu$ branching fraction~\cite{conjugate}, measured by \sbabar~\cite{babar11} and Belle~\cite{belle11}, is very sensitive to contributions from a charged Higgs boson.
The PDG average ${\cal B}(B^+ \ra \tau^+ \nu)=(1.65\pm 0.34) \times 10^{-4}$~\cite{pdg12, HFAG} is larger than the SM prediction of ${\cal B}(B^+ \ra \tau^+ \nu)=(1.2\pm 0.25) \times 10^{-4}$~\cite{silverman}. Even with these high values of ${\cal B}(B^+ \ra \tau^+ \nu)$, we obtain a sizeable allowed $\bar \rho - \bar \eta$ region; there is no conflict with the SM. Belle has now presented a new measurement of ${\cal B}(B^+ \ra \tau^+ \nu)=(0.72^{+0.27}_{-0.25}({\rm stat}) \pm 0.11\ ({\rm sys})) \times 10^{-4}$~\cite{belle12} that reduces the world average to ${\cal B}(B^+ \ra \tau^+ \nu)=(1.14 \pm 0.22) \times 10^{-4}$. Figure~\ref{fig:gfit2} shows our results in the $\bar \rho - \bar \eta$ plane for this world average and Table~\ref{tab:SMfit} summarizes the $95\%~ {\rm CL}$  ranges of unitarity parameters. The inclusion of the present ${\cal B}(B^+ \ra \tau^+ \nu)$ world average  has hardly any impact on the $\bar \rho -\bar \eta$ plane. 


\begin{figure}[h]
\centering
\vskip -8pt
\includegraphics[width=85mm, height=90mm]{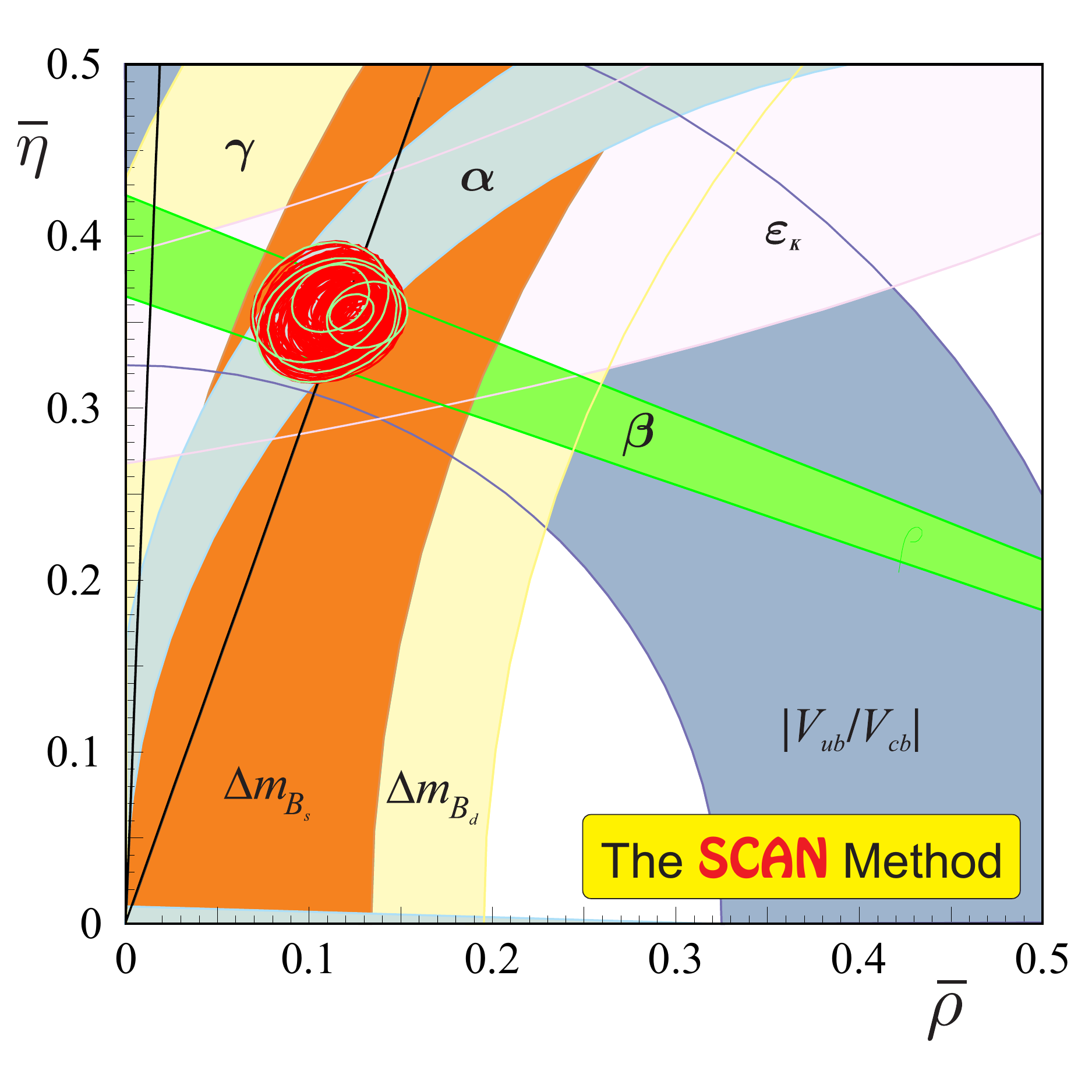}
\vskip -18pt
\caption{Overlay of $95\%~{\rm CL}$ contours in the $\bar \rho -\bar \eta$ plane for fits with 257 measurements without ${\cal B}(B^+  \ra \tau^+ \nu)$. The green ellipses show selected contours of accepted fits.}
\label{fig:fullfit}
\vskip 0.5cm
\end{figure}



\begin{figure}
\centering
\includegraphics[width=77mm]{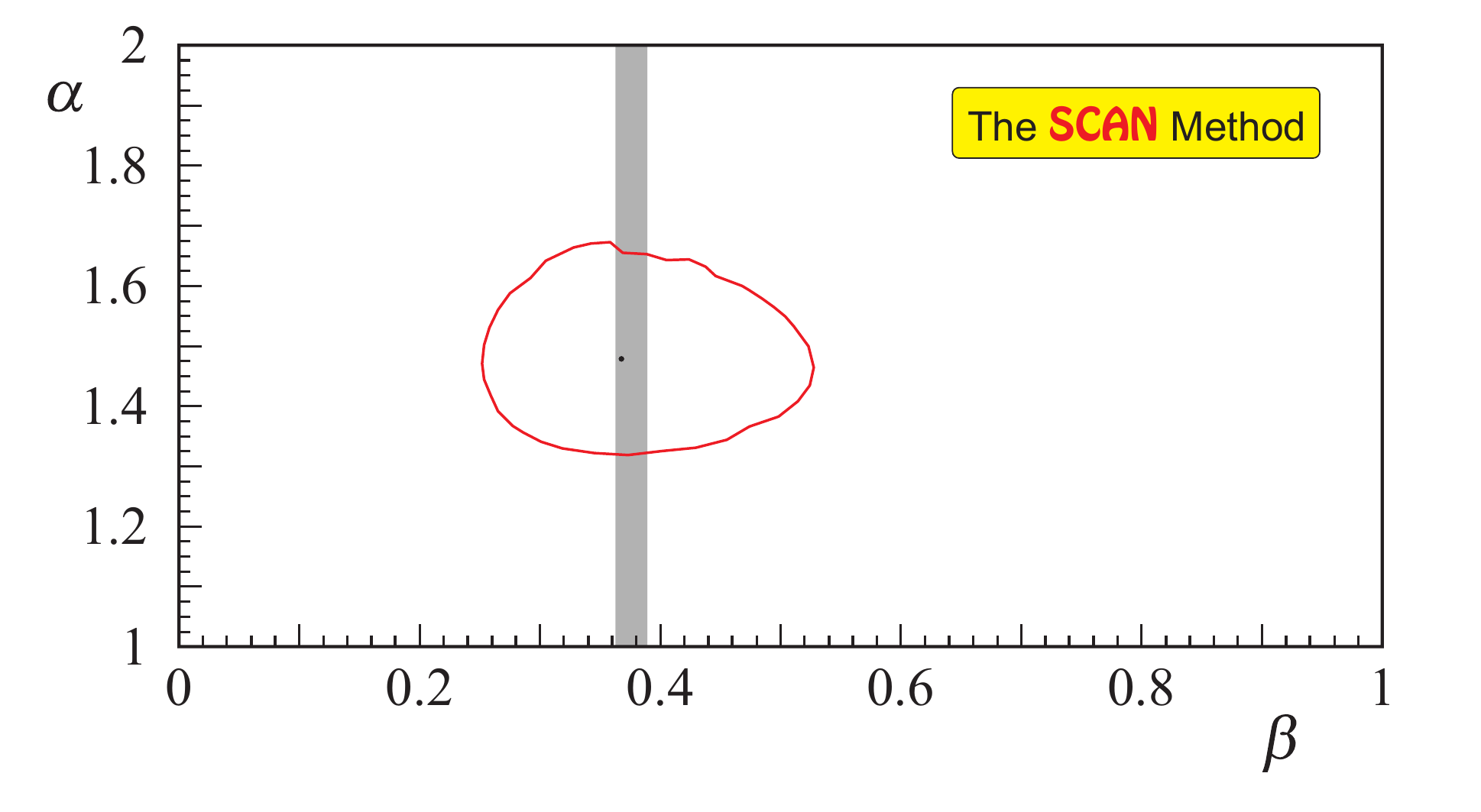} 
\vskip -10pt
\caption{The $95\%~{\rm CL}$ contour in the $\alpha- \beta$ plane from a fit including the $B \ra PP$, $B \ra PV$, $B \ra VV$ and $B \ra P a_1$ modes. The black dot shows the central value from the fit.  The vertical band shows the $68\%~{\rm CL}$ range for $\beta$ obtained from $\sin 2 \beta$ measurements~\cite{HFAG}.}
\label{fig:alpha}
\end{figure}

\subsection{Fit Results using Individual Measurements of Branching Fractions and \CP Asymmetries of Various Decays}
\label{fullfit}

We also perform fits, called type III fits, in which, instead of treating the values of $\alpha$ and $\gamma$ as inputs, we directly include the measurements that determine them.  This allows us to determine the correlations in the extraction of the various unitarity triangle angles from the fit.
We omit ${\cal B}(B^+ \ra \tau^+ \nu)$ in these fits; this has little effect on the results.  

We replace the direct $\alpha$ measurement term in the $\chi^2$ function by all measured branching fractions and \CP asymmetries in $B \ra PP$,  $B \ra PV$, $B \ra VV$ and $B \ra a_1P$ modes in the fit.
Following the Gronau-Rosner approach~\cite{gronau}, we parametrize amplitudes in terms of tree, color-suppressed tree, penguin, singlet penguin, $W$-annihilation/$W$-exchange, electroweak and color-suppressed electroweak diagrams (up to $\lambda^2$ beyond leading order).  Thus for order $\lambda^2$, we consider SU(3) corrections for the tree, color-suppressed tree and penguin diagrams. Since calculations of branching fractions use the $B^+$ and $B^0_d$ lifetimes, we add $\chi^2$ terms for these lifetimes in the fit.

We replace the direct $\gamma$ measurement term in the $\chi^2$ function, by all  measured branching fractions and \CP asymmetries for $B^+  \ra D^{(*)} K^+$ and $B^+ \ra D K^{*+}$ decays analyzed in the Giri-Grossman-Soffer-Zupan (GGSZ)~\cite{ggsz}, Gronau-London-Wyler (GLW)~\cite{glw} and Atwood-Dunietz-Soni (ADS) methods~\cite{ads}. We also include branching fractions and \CP asymmetries for $B^+  \ra D^{(*)} \pi^+$ decays analyzed in the ADS method. 
Here, the observables are calculated in terms of $b \ra c \bar u s (d)$ and $ b \ra u \bar c s (d)$ amplitudes. 
We also include time-dependent \CP asymmetries in $B^0 \ra D^{(*)+} \pi^-$ and $B^0 \ra D^{(*)+} \rho^-$ decays that determine $\sin (2 \beta + \gamma)$. 
Thus, the total number of measurements in the global fits increases to 257;  the fit has 120 parameters.
Figure~\ref{fig:fullfit} shows the $95\%~{\rm CL}$ contours of all accepted fits in the $\bar \rho - \bar \eta$ plane. Table~\ref{tab:SMfit} summarizes the $95\%~{\rm CL}$ ranges for unitarity triangle parameters.
This result shows that the 257 measurements are in good agreement with the SM. This procedure accounts for possible correlations between $\alpha, \gamma$ and the other Wolfenstein parameters ({\it e.g.}, $\beta$). 

\section{Determination of $\alpha$ and $\gamma$}
\label{sec:alphaGamma}

In addition to the global fits of the CKM matrix, we perform separate fits that determine the unitarity triangle angles $\alpha$ and $\gamma$. These results are then used as inputs to the baseline fits. 
They are also used to investigate the correlation with the angle $\beta$.

\subsection{Determination of $\alpha$}

For the $\alpha$ determination we combine all measured branching fractions and \CP asymmetries in $B \ra PP$,  $B \ra PV$ and $B \ra VV$ modes. 
We first perform separate fits for each class of decays. 
As in Section~IV
we parameterize the observables in terms of amplitudes, following the Gronau-Rosner method. We include $f_{B_s}$ and $\xi_f$
in the fit, but we do not scan over them, since these parameters appear only in the $W$-exchange and $W$-annihilation diagrams that are at order $\lambda^2$ with respect to the tree diagram; any variation in these parameters is absorbed by adjusting the magnitude of the $W$-annihilation/$W$-exchange diagrams, leaving $\alpha$ and the remaining parameters unchanged.  Thus, we perform a single fit for
each class of decays and plot $95\%$~{\rm CL} $\alpha-\beta$ contours. 
Table~\ref{tab:alpha} shows  the central values for $\alpha$ and $\beta$ with uncertainties obtained by changing
$\Delta \chi^2$ by one for all fits. In addition, we list the correlation between $\alpha$ and $\beta$ and the fit probability. The correlation coefficients vary between -18\% and 5\%.

We next perform a combined fit of all measurements in $B \ra PP, B \ra PV$ and $B \ra VV$ and $B \ra Pa_1$ modes to extract $\alpha$. We use 185 measurements~\cite{HFAG} to determine 96 parameters. 
Figure~\ref{fig:alpha} shows the $95\%~{\rm CL}$  contour in the $\alpha-\beta$ plane, which encompasses the world average $ \beta = (21.4 \pm0.8)^o$ measured using $b \ra c \bar cs$ modes.
The fit probability is $P(\chi^2) = 38.6\%$. The correlation coefficient is about -4\%. 
These results show that all measurements  in $B \ra PP, B \ra PV$ and $B \ra VV$ and $B \ra Pa_1$ modes are consistent with the SM description and no new physics amplitudes are required. 

\subsection{Determination of $\gamma$}

\begin{table*}[tbp!]
\centering
\caption{Measurements of $\alpha$, $\beta$ and $\gamma$ from fits of branching fractions and \CP asymmetries in $B \ra PP$, $B \ra PV$,  $B \ra VV$ decays and in a combination of all modes combined plus $B \ra P a_1$ decays ($\alpha$) and $B \ra D^{(*)} K (\pi) + B \ra D K^* (\rho)$ decays ($\gamma$).}\medskip
{
\begin{tabular}{|l|c|c|c|c|c|}
\hline \hline 
 & $B \ra PP$ & $B \ra PV$ & $B \ra VV$ & $B$ modes combined & $B \ra D^{(*)} K (\pi) + B \ra D K^* (\rho)$ \T\B \\ \hline
$\ \alpha~ [^\circ]$   & $ 85.9^{+3.0}_{-2.7} $ &  $ 82.4^{+4.1}_{-4.3}$ & $ 83.8^{+5.5}_{-5.6}$ & $ 84.7^{+2.1}_{-2.1}$   & - \TT\B\\
$\ \gamma~[^\circ]$ &     -                                  &                  -                      &              -       & -  & $79.6^{+4.1}_{-4.2}$ \T\B \\                  
 $\ \beta~[^\circ]$     & $ 20.8^{+2.1}_{-1.9}$ & $ 20.5^{+3.6}_{-3.4}$ & $ 24.3^{+6.4}_{-4.9}$ &  $21.1^{+1.6}_{-1.6}$ &  $22.8^{+7.7}_{-2.1}$ \T\B \\
$\alpha - \beta$ correlation & 0.052 & -0.182 & -0.151 & -0.035 & - \T\B \\
$\gamma - \beta$ correlation & - & - & - & -& -0.194 \T\B \\
p-value & 0.50  &  0.369 & 0.248 & 0.386 & 0.051-0.071 \T\B \\
\hline \hline
\end{tabular}
}
\label{tab:alpha}
\end{table*}

For the $\gamma$ determination, we use branching fraction and \CP asymmetries of $B^+  \ra D^{(*)} K^+$ and $B^+ \ra D K^{*+}$ decays analyzed in the GGSZ~\cite{ggsz}, GLW~\cite{glw}  and ADS~\cite{ads} methods.  We also include branching fractions and \CP asymmetries of $B^+  \ra D^{(*)} \pi^+$ decays analyzed in the ADS method and time-dependent \CP asymmetries in $B^0 \ra D^{(*)+} \pi^-$ and $B^0 \ra D^{(*)+} \rho^-$ decays that determine $\sin (2 \beta + \gamma)$. We separate the CKM factors, $|V_{us} V_{ub}^*|/ |V_{cs}V_{cb}^*|$ and $|V_{ud} V_{ub}^*|/|V_{cd} V_{cb}^*|$  from the ratio of $b\ra u$ to $b \ra c$ amplitudes. We include the ratios 
$|V_{us}/V_{ud}|$ and $|V_{ub}/V_{cb}|$ in the fit, scanning over $|V_{ub}|$ and $|V_{cb}|$.  
Since the predictions contain a product of CKM factors and ratios of amplitudes, it is necessary to constrain the CKM factors in the fit to obtain sensible values for the amplitude ratios and CKM factors. 

\begin{figure}[h]
\centering
\hspace{-8mm}
\includegraphics[width=74mm]{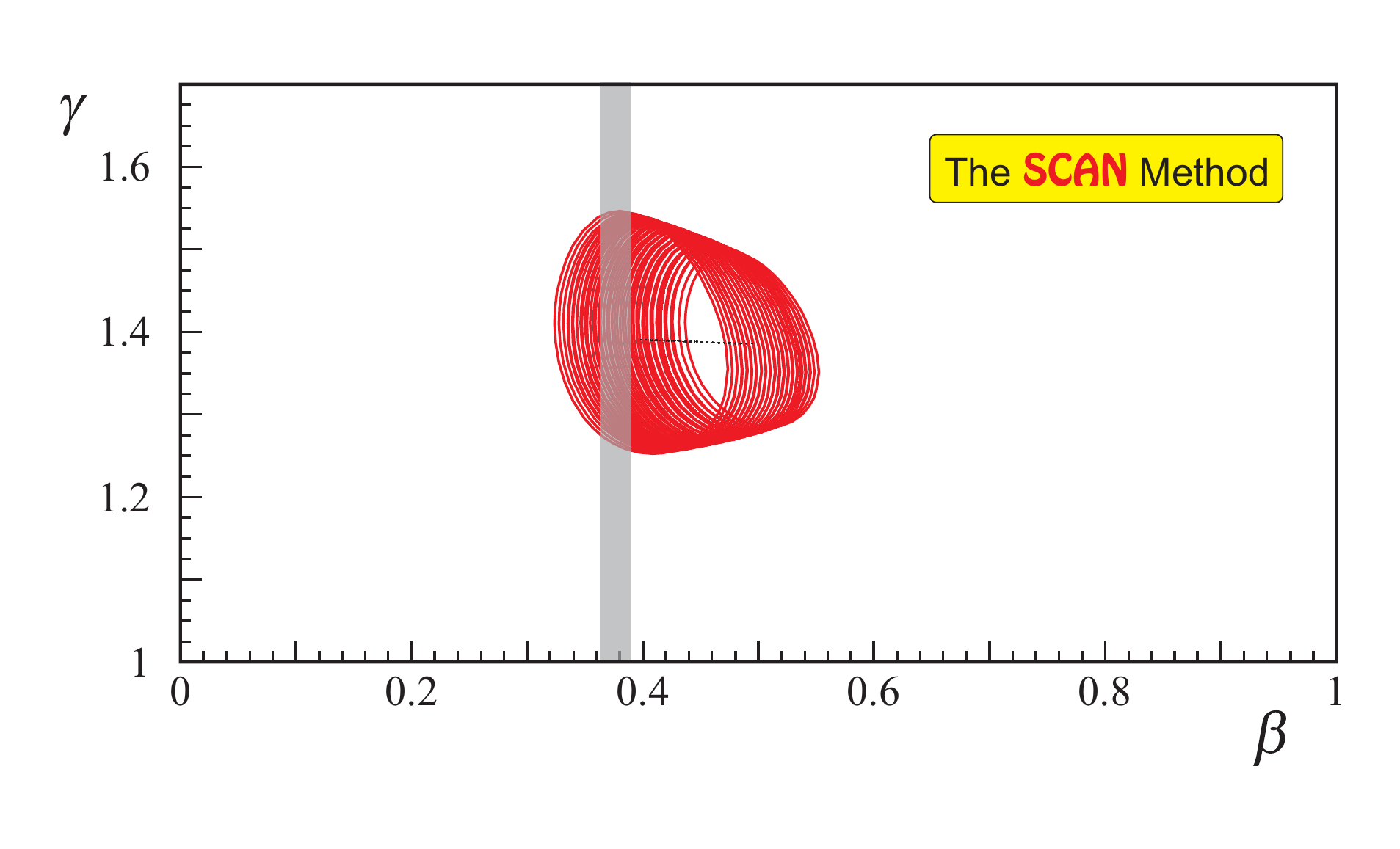}
\vskip -10pt
\caption{Overlay of $95\%~{\rm CL}$ contours in the $\gamma- \beta$ plane for fits including the $B  \ra DK^{(*)}, DK^*, D^{(*)} \pi, D \rho$ modes. The vertical band shows the $68\%~{\rm CL}~\beta$ region obtained from $\sin 2 \beta$ measurements~\cite{HFAG}.}
\label{fig:gamma}
\end{figure}

We use 56 measurements~\cite{HFAG} to extract 19 fit parameters, scanning over the constraint $|V_{ub}|/|V_{cb}|$. The probabilities for
these fits range between $P(\chi^2) = 5.1\%$ and $P(\chi^2) = 7.1\%$. Figure~\ref{fig:gamma} shows the resulting  contours at $95\% ~{\rm CL}$ in the $\gamma -\beta$ plane. Table~\ref{tab:alpha} lists the fit results. Again, these 56 modes are well-described within the SM and no new physics amplitudes are required. The $\gamma$--$\beta$ correlation coefficient is $-19$\%.

\section{Conclusion}
\label{sec:conclusion}

The three fitting approaches: CKMfitter, \UTfit and the scan method yield similar central values for $\bar \rho$  and $\bar \eta$ when presented with identical inputs.
However, the allowed region in the $\bar \rho-  \bar \eta$ plane is substantially larger in the 
scan method. This may be expected from the difference with the Bayesian methodology
of \UTfit$\!\!$, in which prior distributions are assigned for the theoretical uncertainties.
The differences with
CKMfitter are more subtle, as both are frequentist approaches. In fact, the
larger scan method intervals is not a given; with different measurement values the
situation could reverse in the comparison with CKMfitter.  This comparison points out the importance of
methodology in forming conclusions, and hence of examining the problem with multiple approaches.

Using the scan method, we find no tension with the SM even
when we include the current PDG value of ${\cal B}(B^+ \ra \tau^+
\nu)=(1.65\pm 0.34) \times 10^{-4}$; the scan yields global fits consistent
with the SM at $95\%~{\rm CL}$. When we include the recent Belle result, the ${\cal B}(B^+ \ra \tau^+ \nu)$ branching fraction has hardly any impact on the $\bar \rho - \bar \eta$ plane.

Our global fit allows us to determine and incorporate the correlation between $\alpha$ and $\beta$, as well as between $\gamma$ and $\beta$.
Using all measured branching fractions and \CP asymmetries of $B \ra PP$, $B \ra PV$, $B \ra VV$, $B \ra a_1 P$ modes and $ B^+\ra D^{(*)} K^+, B \ra D K^{*+}$ modes that are sensitive to $\alpha$ and  $\gamma$, respectively, we observe small changes in the
allowed region in the $\bar \rho - \bar \eta$ plane. 
From separate fits of branching fractions and \CP asymmetries in these modes, we determine $\alpha-\beta$ and $\gamma-\beta$ contours. Though $\alpha$ measurements agree with each other and $\beta$ results are consistent with $\sin 2\beta$ from $b \ra c \bar cs$ modes, some correlation among Wolfenstein parameters in the different measurements is observed.
The values of $\alpha$, determined from a fit to all measured branching fractions and \CP asymmetries of $B \ra PP$, $B \ra PV$, $B \ra VV$ and $B \ra P a_1$ modes, and $\gamma$, extracted from a fit to  $ B^+\ra D^{(*)} K^+,  B \ra D K^{*+}$, $B \ra D^{(*)} \pi$ and $B^0 \ra D^+ \rho^-$ modes, agree with the SM expectations.

\section*{Acknowledgments}

This work is supported in part by the U.\ S.\ Department of Energy under Grant DE-FG02-92-ER40701
and by the NFR (Norway).

\vfil
 
\break

\bigskip
\section*{Appendix}

In the baseline fits, the $\chi^2$  function (Eq.~\ref{eq:chisq}) includes 23 or 22 terms, depending on whether ${\cal B}(B\to \tau\nu)$ is included or not. In this Appendix, we describe the dependence of the predicted values 
used in the $\chi^2$ expression on the quantities $\bar \rho, \bar \eta, p_i$, and $t_j$.

The CKM matrix elements are parameterized in terms of Wolfenstein parameters $\bar \rho$, $\bar \eta$, $A$ and $\lambda$ up to order ${\cal O}(\lambda^9)$~\cite{wolf, buras94, ahn}:


\begin{widetext}
\begin{align}
V_{ud} =& 1- \frac{1}{2} \lambda^2 - \frac{1}{8} \lambda^4 -\frac{1}{16} \lambda^6 \bigl(1+ 8 A^2 (\rho^2 +\eta^2) \bigr) -\frac{1}{128} \lambda^8 \bigl(5- 32 A^2 (\rho^2 + \eta^2) \bigr), \nonumber \\
V_{us} =& \lambda \bigl( 1 -\frac{1}{2} A^2 \lambda^6 (\rho^2 + \eta^2) \bigr), \nonumber \\
V_{ub} =& A \lambda^3 (\rho - i \eta), \nonumber \\
V_{cd}=& - \lambda \Bigl (1 - \frac{1}{2} A^2 \lambda^4 \bigl( 1-2(\rho + i \eta) \bigr ) - \frac{1}{2} A^2 \lambda^6 (\rho + i \eta) \Bigr ), \nonumber \\
V_{cs}=& 1- \frac{1}{2} \lambda^2 - \frac{1}{8} \lambda^4 \bigl( 1+4A^2 \bigr)  - \frac{1}{16} \lambda^6 \bigl(1- 4 A^2 +16 A^2 (\rho +i \eta)  \bigr) 
 -\frac{1}{128} \lambda^8 \bigl(5- 8 A^2 +16 A^4 \bigr), \nonumber \\
 V_{cb}=& A \lambda^2 (1-\frac{1}{2} A^2 \lambda^6  \bigl( \rho^2 + \eta^2 \bigr)), \nonumber \\
 V_{td}=& A \lambda^3 \bigl (1 -\rho -i \eta \bigr) + \frac{1}{2} A \lambda^5 \bigl( \rho +i \eta \bigr) + \frac{1}{8} A \lambda^7 \bigl( 1+4A^2 \bigr)  \bigl( \rho +i \eta \bigr), \nonumber \\
 V_{ts} =& -A \lambda^2 \Bigl( 1-  \frac{1}{2} \lambda^2  \bigl( 1-2(\rho + i \eta) \bigr ) - \frac{1}{8} \lambda^4 - \frac{1}{16} \lambda^6 \bigl( 1+8A^2 (\rho + i \eta) \bigr)   \Bigr), \nonumber \\
 V_{tb} =& 1-  \frac{1}{2} A^2  \lambda^4 -  \frac{1}{2} A^2  \lambda^6 (\rho^2 + \eta^2) -\frac{1}{8} A^4 \lambda^8. 
\end{align}
\end{widetext}

The  Unitarity Triangle angles $ \beta$, $\alpha$ and  $ \gamma$ are implemented by

\begin{eqnarray}
\sin 2 \beta &=& \frac{2 \bar \eta (1-\bar \rho)} {(1-\bar \rho)^2 +\bar \eta^2}, \nonumber \\
\tan \alpha &=& \frac{\bar \eta} {\bar \eta^2 +\bar \rho(\bar \rho -1)},  \nonumber\\
\tan \gamma&=& \frac{\bar \eta}{\bar \rho}.
\end{eqnarray} 

The oscillation frequencies for $B^0_d \bar B^0_d$ and $B^0_s \bar B^0_s$ mixing are 
computed according to:

\begin{eqnarray}
\Delta m_{B_d} &=& \frac{ G^2_F}{6 \pi^2} \eta_B m_{B_d} \frac{f^2_{B_s}}{\xi_f^2} \frac{B_{B_s}}{\xi_b} m_W^2 S(x_t) |V_{td}V^*_{tb}|^2, \nonumber \\
\Delta m_{B_s} &=& \frac{ G^2_F}{6 \pi^2} \eta_B m_{B_s} f^2_{B_s} B_{B_s}  m_W^2 S(x_t) |V_{ts}V^*_{tb}|^2, 
\end{eqnarray}
\noindent
where $G_F$ is the Fermi constant, $\eta_B$ is a QCD correction, $m_W$ is the $W$ mass, $S(x_t)$ is the Inami-Lim function~\cite{inami} and $x_t= \overline{m}_t^2/m^2_W$ where the top quark mass is calculated in the $ \rm \overline{MS}$ scheme. We have expressed the $B^0_d$ decay constant and bag parameters in terms of the $B^0_s$ decay constant and bag parameters and their ratios $\xi_f$ and $\xi_b$ since the latter have smaller uncertainties.
The explicit relation between $\overline{m}_t(m_t)$ and $m_t^{\rm pole}$ is given by
\begin{equation}
\overline{m}_t (m_t)=m_t^{\rm pole} \bigl(1-\frac{4}{3} \bigl ( \frac{\alpha_s}{\pi} \bigr ) - 9.1253 \bigl ( \frac{\alpha_s}{\pi} \bigr )^2 -80.4045 \bigl ( \frac{\alpha_s}{\pi} \bigr )^3 \bigr),
\end{equation}
where $\alpha_s(\overline{m}_t) =0.1068 \pm 0.0018$ is calculated in the $\overline{MS} $ scheme for six quark flavors at the scale of the pole mass~\cite{buras97, ckmfitter}.

\CP violation in the $K^0 \bar K^0$ system is represented by the parameter $\epsilon_K$. In the SM, this is proportional to the off-diagonal matrix element of the mixing matrix divided by the $K^0_L- K^0_S$ mass difference $\Delta m_K$ yielding~\cite{buchalla95}
 
\begin{widetext}
\begin{equation}
\epsilon_K = C_\epsilon \kappa_\epsilon \exp{i \phi_\epsilon} B_K \Bigl( {\rm Im}[ (V_{cs}V^*_{cd})^2] \eta_{cc} S(x_c) + {\rm Im}[ (V_{ts}V^*_{td})^2] \eta_{tt} S(x_t) +
 2{\rm Im}[ V_{cs}V^*_{cd} V_{ts}V^*_{td}] \eta_{ct} S(x_c, x_t) \Bigr),
\end{equation}
\end{widetext}
where $\kappa_\epsilon = 0.94 \pm 0.02$~\cite{buras10}, $\phi_\epsilon=(43.5 \pm 0.7)^\circ$, $x_c= \overline{m}_c^2/m^2_W$, $\overline{m}_c$ is the charm quark mass in the $ \rm \overline{MS}$ scheme. The constant is given by
\begin{equation}
C_\epsilon = \frac{G^2_F f^2_K m_K m^2_W}{12 \pi^2 \sqrt{2} \Delta m_K},
\end{equation}
where $m_K$ and $f_K$ are the kaon mass and kaon decay constant, respectively. The contribution from the decay rate difference has been neglected. We use NLO calculations of the QCD parameters;  $\eta_{cc}$ and $\eta_{ct}$ calculations at NNLO~\cite{brod10, brod12} have now been done.

The $B^+ \ra \tau^+ \nu$ branching fraction is given by

\begin{equation}
{\cal B} (B^+ \ra \tau^+ \nu) = \frac{G^2_F}{8 \pi} m_{B^+} m^2_\tau \Bigl(   1- \frac{m^2_\tau}{m^2_{B^+}} \Bigr) \frac{f^2_{B_s}}{\xi_f^2} |V_{ub}|^2 \tau_{B^+},
\end{equation}
 where $\tau_{B^+}$ is the $B^+$ lifetime, $m_{B^+}$ is the $B^+$ mass and $m_{\tau}$ is the $\tau^+$ mass,

We also add Gaussian terms in the $\chi^2$ function for the quark masses $m_t ^{\rm pole}$ and  $\overline{m}_c(m_c)$, meson messes $m_{B^0_d}$ and  $m_{B^0_s}$, and the Gaussian parts of the QCD parameters $B_K$, $f_{B_s}$, $\xi_f$, $B_{B_s}$, and $\xi_b$.

\vfil

\end{document}